\pdfoutput=1
\documentclass[conference]{IEEEtran}
\IEEEoverridecommandlockouts
\usepackage{cite}
\usepackage{amsmath,amssymb,amsfonts}
\usepackage{algorithmic}
\usepackage{graphicx}
\usepackage{textcomp}
\usepackage{xcolor}
\def\BibTeX{{\rm B\kern-.05em{\sc i\kern-.025em b}\kern-.08em
    T\kern-.1667em\lower.7ex\hbox{E}\kern-.125emX}}

\usepackage{soul}
\usepackage{tikz}
\usetikzlibrary{calc}

\makeatletter
\newif\if@anonymize

\@anonymizefalse  

\if@anonymize
  \newcommand{\highlight@DoHighlight}{
    \fill [outer sep = -15pt, inner sep = 0pt, color=black]
          ($(begin highlight)+(0,8pt)$) rectangle ($(end highlight)+(0,-3pt)$) ;
  }

  \newcommand{\highlight@BeginHighlight}{
    \coordinate (begin highlight) at (0,0) ;
  }

  \newcommand{\highlight@EndHighlight}{
    \coordinate (end highlight) at (0,0) ;
  }

  \newdimen\highlight@previous
  \newdimen\highlight@current
  \newlength{\item@width}

  \DeclareRobustCommand*\anonymize{%
    \SOUL@setup
    \def\SOUL@preamble{%
      \begin{tikzpicture}[overlay, remember picture]
        \highlight@BeginHighlight
        \highlight@EndHighlight
      \end{tikzpicture}%
    }%
    \def\SOUL@postamble{%
      \begin{tikzpicture}[overlay, remember picture]
        \highlight@EndHighlight
        \highlight@DoHighlight
      \end{tikzpicture}%
    }%
    \def\SOUL@everyhyphen{%
      \discretionary{%
        \SOUL@setkern\SOUL@hyphkern
        \SOUL@sethyphenchar
        \tikz[overlay, remember picture] \highlight@EndHighlight ;%
      }{%
      }{%
        \SOUL@setkern\SOUL@charkern
      }%
    }%
    \def\SOUL@everyexhyphen##1{%
      \SOUL@setkern\SOUL@hyphkern
      \settowidth{\item@width}{##1}%
      \makebox[\item@width]{}%
      \discretionary{%
        \tikz[overlay, remember picture] \highlight@EndHighlight ;%
      }{%
      }{%
        \SOUL@setkern\SOUL@charkern
      }%
    }%
    \def\SOUL@everysyllable{%
      \begin{tikzpicture}[overlay, remember picture]
        \path let \p0 = (begin highlight), \p1 = (0,0) in \pgfextra
          \global\highlight@previous=\y0
          \global\highlight@current =\y1
        \endpgfextra (0,0) ;
        \ifdim\highlight@current < \highlight@previous
          \highlight@DoHighlight
          \highlight@BeginHighlight
        \fi
      \end{tikzpicture}%
      \settowidth{\item@width}{\the\SOUL@syllable}%
      \makebox[\item@width]{}%
      \tikz[overlay, remember picture] \highlight@EndHighlight ;%
    }%
    \SOUL@
  }
\else
  \newcommand{\anonymize}[1]{#1}
\fi
\makeatother

\begin{document}

\title{A Code Injection Method for Rapid Docker Image Building}

\author{\IEEEauthorblockN{\anonymize{Yujing Wang}}
\IEEEauthorblockA{\textit{\anonymize{Department of Mechanical and Mechatronics Engineering}} \\
\textit{\anonymize{University of waterloo}}\\
\anonymize{Waterloo On, Canada} \\
\anonymize{yj9wang@edu.uwaterloo.ca}} \and
\IEEEauthorblockN{\anonymize{Qinyang Bao}}
\IEEEauthorblockA{\textit{\anonymize{Department of Mechanical and Mechatronics Engineering}} \\
\textit{\anonymize{University of waterloo}}\\
\anonymize{Waterloo On, Canada} \\ 
\anonymize{q7bao@edu.uwaterloo.ca}}}
\maketitle
\begin{abstract}
Docker images are composed of multiple layers, each of which contains a set of instructions, and an archive of files. Layers allow Docker to separate a large build task into smaller ones, such that when a part of the program is changed, only the corresponding layer needs to be changed. Yet the current implementation has major inefficiencies that make the rebuilding of an image unnecessarily slow when changes in bottom layers are required: uneven content distribution amongst layers, the need to rebuild an entire layer during update, and the rebuild fall-throughs in many cases. In this paper, we propose a code injection method that overcomes these inefficiencies by targeting only the changed layer and then bypassing the layer's content checksum. This process is developed specifically for an interpreted language such as Python, where changes can be detected explicitly via text diff tools and run as-is without compilation. We then demonstrate that this method can accelerate the rebuild time, effectively reducing the O(n) where n = size of layer rebuild time to O(1). Whereas for compiled languages, literal code injection cannot guarantee integrity in compiled machine code. Expanding on the same code injection principle, multi-layer targeted code injection will be addressed in a future discussion. \end{abstract}
\section{Introduction}
\begin{figure}[htbp]\centerline{\includegraphics[width=1\linewidth,scale=0.5]{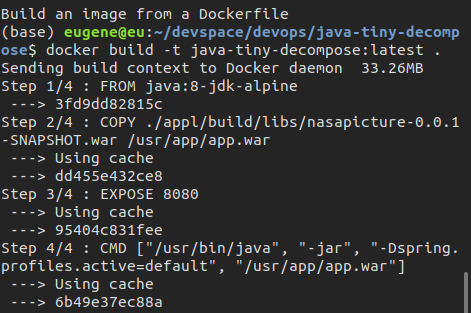}}\caption{Docker Layer and UUID}\label{figure1} \end{figure}

When deploying an application on a single machine can no longer match its expanding usage, developers split it into microservices, and deploy them on multiple machines. They package modules along with its dependencies in an image to guarantee consistency across platforms and deploy these image in containers which are running instances of their images. For system-level processes, developers use a Linux Container (LXC), which uses Linux namespace, a kernel feature that partitions a set of resources for a set of processes exclusively, along with control group (cgroup), a kernel feature that limits and isolate machine resource usage \cite{bibilxc}. On an application level, Developers use Docker that builds an image from instructions given in a Dockerfile. Docker reads each run line of instruction, which is made up of an "Instruction" and its "Arguments", executes it, and stores results in files in an image layer \cite{bibidockerref}. Each layer generated will be assigned a permanent UUID in SHA-256; each revision of a layer will be given a checksum in SHA-256. These values are stored in an image's manifest, so Docker knows which image a layer belongs to and which revision it should use. If a developer changes the content of a layer, the layer's ID remains the same, but its checksum varies. Fig.~\ref{figure1} shows docker building an image layer with id b248b9e23166 from a command "FROM python:alpine" using cache. Notice that after each build, Docker informs the user of each layer's ID. To examine what command each layer corresponds to, the user can run "docker history image:tag". By default, all layers are stored in "/var/lib/docker/overlay2/". A developer can export the image by "docker save image:tag $>$ file.tar" and load it by "docker load $<$ file.tar". The layer by layer architecture speeds up the image building process and makes it more flexible. When Docker creates a new image, it first searches in a registry to find if an exact image layer has already existed. If Docker finds such a layer, it keeps a reference to the layer and proceeds to the next layer even if the layer is from a different image. This process is called "layer deduplication" and is used most commonly for common base layers such as "From ubuntu". When building a new revision of an image, Docker uses layers built in previous revisions for unchanged layers. This process is referred to as "caching", and is used when building a new version of the image where only a small portion of the code is updated. 
\subsection{Docker Layer Caching (DLC) Mechanism}
When the developer runs build command, Docker looks at the following criteria to determine whether or not to use the cache \cite{bibidockerref}:
\begin{enumerate}
    \item Use the parent image as the starting point, pull out its manifest, and examine checksums and UUID of its child images to see if the new build is identical as the existing image. If true, skip building.
    \item Examine a new version of Dockerfile to see if the instruction has been added, removed, or altered, if true, remove or alter the corresponding layer.
    \item For 'ADD', 'COPY' that are altered, compute the checksum of updated files, compare it against existing files if the checksum does not match, 'COPY', or 'ADD' new files to build. The checksum uses the SHA-256 hash algorithm and last modified and last accessed time are not taken into consideration. If they match, use the cache.
    \item For operation commands including but not limited to "RUN", "CMD", "ENTRYPOINT", Docker checks the literal message without checking the corresponding files. for example, for the command "RUN apt install ubuntu" the literal command is checked instead of comparing every file of Ubuntu in the new version against the old version.
\end{enumerate}

In this paper, we will discuss the inefficiency of Docker's deduplication and caching process that resulted in unnecessary layer rebuilds. We propose a code injection method that bypasses Docker's rebuild mechanism injects changes directly into the targeted layer(s) and reconfigures layers' checksum to bypass integration tests. This method effectively eliminates the image layer rebuild in programming languages where the code is directly interpreted and only that content has been changed. The technique effectively transforms the image layer rebuild process from linear O(n) time to constant O(1) time for the aforementioned scenario. To demonstrate the result, we conducted an experiment executed across three different machines, each for a total of 100 trials. In the test cases, the method was able to shorten layer rebuild speed significantly.    

\section{Inefficiency in Docker's Deduplication and Caching Mechanism}
When building an image, if a layer already exists in Docker's registry, Docker adds the layer's UUID into the image's manifest so that it becomes a part of the new image. Docker then proceeds to build other layers. If the code concerning one layer has been altered, Docker builds a new layer and changes the layer pointer in the new image to point to the newly created layer. The old layer is preserved and is still referenced by the old image. The old layer can be deleted if only all references to it have been removed. By this mechanism, no matter how small of a change, a layer always has to be rebuilt. Although an image being broken down into layers already helps modularizing an image such that Docker does not need to rebuild an entire image each time a new version is tagged. Having to perform the rebuilt of a full layer is nevertheless wasteful and unnecessary in many situations. 

\subsection{Uneven Distribution of Layer Content}
When Docker builds an image, the content is not evenly distributed across all layers. A configuration layer, which begins with "ENV", "ENTRYPOINT", "CMD", and "Label" instructions, describe a state of the application and contains no content files. A content layers, built by "ADD", "COPY", "RUN" or "FROM" instructions, contains files generated as a result of the instruction. There is a significant size disparity among these two types of layers. When a layer is small, building the layer itself incurs significantly higher overheads than building the items in these layers. When small layers stack together, the compile-time accumulates. 

\subsection{Rebuilding a Large Layer for Small Change}
When updating a large Layer, the system needs to rebuild the entire layer because a layer is the smallest Docker addressable image component. If a line of little significance in a enormous project is updated, the entire layer that imports the project must be updated. Take the extreme case of changing one line of comment in a 20Gi image as an example. Changing a line of comment does not change any operation logic; however, as Docker will detect such a change, and subsequently, rebuild the entire layer, which costs about 20 minutes as the experiment we performed below shows. 

\subsection{ Layer Fall-through}
\begin{figure}[htbp]\centerline{\includegraphics[width=\linewidth]{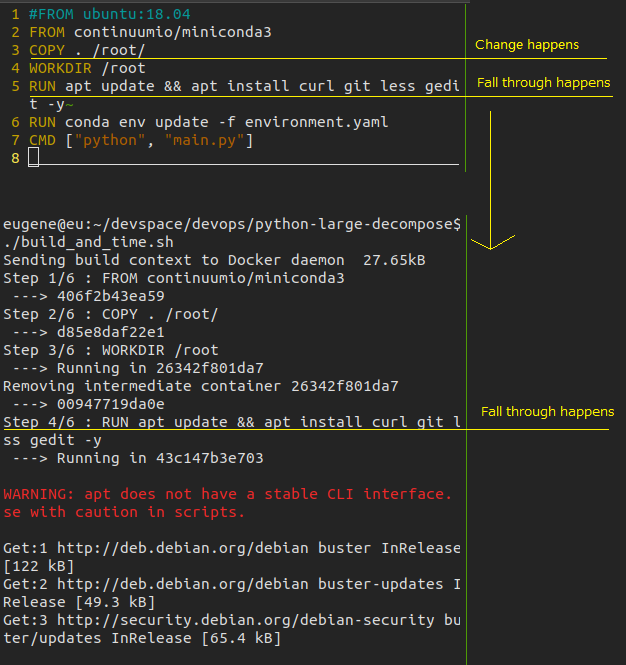}}\caption{Changes in step 2 (left) resulted in a fall through on step 4 onwards(right)}\label{figfallthrough} \end{figure}

Most projects import code early in a Dockerfile. In the example of Fig~\ref{figfallthrough}, the Dockerfile command "COPY . ." at line 2 copies all files in the current directory into an image. If a developer then alters files in the project's scope after the initial build, Docker will detect a change in layer 2, it rebuilds every layer starting from layer 2 to the end. This fall-though occurs as the lower layers depends on top layers a cotent-wise change "ADD" or "COPY" change leads to redo of all operation-wise layers "RUN" or "ENV" even if they are unrelated.

Compounding unevenly distributed layers, rebuilding an entire layer and layer falls-through, one would get a significant delay in the image build process. Coupled with the complex deployment scenarios, building images for large applications becomes extremely time consuming. Updating a layer larger than 20GiB may take more than 10 minutes. Also, the modern software development process encourages a build after each small incremental change such that a pipeline checks for whether the new feature update works as intended. This becomes problematic when we have a high demand for builds but a low throughput of build runtime, which is clogged up by long build time. 

\section{Process}
We propose a methodology that injects newly edited code into the existing docker image layer while bypassing the SHA-256 checksum rule to enable rapid docker image building without rebuilding an entire layer. This reduces O(n) linear runtime required for updating an image where n = size of the layer to a constant O(1) runtime for interpreted languages.

\subsection{Code Injection}
Interpreted languages such as Python, PHP, Perl, Ruby, and Javascript are written in literal text and run as it is. The code is picked up by an interpreter pre-installed on the system. Interpreted languages typically perform slower than compiled languages, which have source code compiled to machine code before execution. The advent of in just-in-time(JIT) compiler makes interpreted language such as Python usable for industry applications.
\begin{figure}[htbp]\centerline{\includegraphics[width=\linewidth]{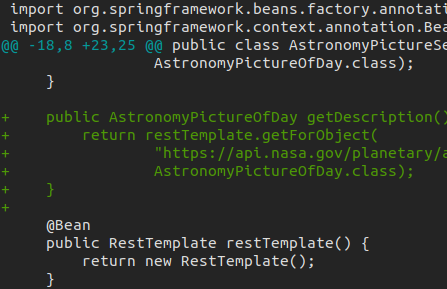}}\caption{Using 'diff' to check changes between old and new revision}\label{figcomparator} \end{figure} When building a Docker image for an interpreted language code base, newly edited code can be compared side by side against the original code to identify where the changes occur, as Fig.~\ref{figcomparator} shows. If the developer has made changes that only affect one layer, it is unnecessary to build a new layer and all layers after it. 

In the code injection method, we first use Docker's system mechanism to pull out the old image and proceed down the Dockerfile line by line to check which layer has been changed. Once such a changed layer is detected, determine what type of change this is: 
\begin{enumerate}
    \item A content change "ADD" or "COPY"
    \item A configuration change
\end{enumerate}
If it is a type 1 change, decompose the original layer and obtain a collection of files. This can be done in 2 ways. The explicit way is to export the image with docker api "docker save image\_name:image\_tag $>$ archive\_name.tar". Docker outputs a bundled archive of the specified image, containing the image's manifest and its layers. Each folder of these layers contains a layer.tar, manifest, and a JSON. The content is stored in the layer.tar file and put into a directory named after its UUID. Examining this bundle, developers will be able to compare files from an existing image with the files in the current directory. After the change is determined, inject the new code into the files in the image, and save changes. The implicit way is to look at an image's manifest and the 'config'.json file. Table~\ref{tab:composition} shows the composition of image and layers for implicit decomposition. The manifest details an array of IDs of all its layer, and the way they are organized, whereas the "config".json contains each layer's checksum, update trace, version, and instruction of the image.  By default, all layers are stored in "/var/lib/docker/overlay2/'layer\_id'/". Knowing these changes can be made to the layer directly without having to export the image or import the image. Removing an intermediate stage, decomposing implicitly is much faster than explicitly, as the experiment shows below. 
\begin{table}[htbp]
\label{tab:composition}\caption{Composition of a Image and its layer}
\begin{center}\begin{tabular}{|c|c|c|c|c|}
\hline
\textbf{Item} & File & Content \\ \hline
 & manifest.json & config pointer, RepoTags, list of layer pointers \\ \cline{2-3}
Image & repositories & repository and pointer to latest layer \\ \cline{2-3}
 & 'config'.json & image config, array of layers' config ie: \\ 
 & & arch, version, layer-checksum, instruction ...\\ \hline
 & version & version of this layer \\ \cline{2-3} 
Layer & layer.tar & archive of all files generated at this layer \\ \cline{2-3} 
 & json & layer specific config: id, version-sha, \\
 & & layer-checksum, env, isEmptyLayer, etc.\\  \hline
\end{tabular}\end{center}\end{table}

\subsection{Checksum Bypass}
Changing the content of layer.tar results in a different SHA-256 checksum. Docker uses the SHA-256 algorithm to detect an update or a corruption. The SHA-256 algorithm is a cryptography hash function and used in the digital certificate as well as in data integrity \cite{bibisha}. With a given data of any length, the SHA-256 algorithm pads the data so that the length is a multiple of 512 bits \cite{bibishadetail}. Then split the data into chunks each of 512 bits long: $M^{(1)}, M^{(2)},..., M^{(n)}$. Use $H^{0}$ as a fixed initial hash value base, sequentially compute \begin{equation}
    H^{i} = H^{(i-1)}+C_{M^{(i)}}(H^{(i-1)}),
\end{equation}
where C is the SHA-256 compression function and + is word-wise $2^{32}$ addition, and $H^{(N)}$ is the hash of M. The eventual output of SHA-256 is always a 256 bits hash as its name suggests. This gives a total possibility of $2^{256}$ or roughly $1.16E+77$ combinations. Having such an enormous amount of combinations, SHA-256 is extremely unlikely to run into collision, and none have been reported so far.

Yet in our case, we do not need to break the SHA-256 uses to check the integrity of image layers. Given that we have the id of the layer to be changed and the original checksum, we can search for all occurrences of the original checksum in image's "config".json file. Then inject code, and compute the checksum of the new layer using Linux built in functions: "SHA-256sum $file\_name$". Replace the previously located old image hash with the new ones. This way we update both the key and the lock SHA-256 to bypass integrity test which was put in place to ensure no corruption in the layers. On the other hand, if a type 2 configuration change is detected, let Docker perform the update since a configuration layer is an "empty layer" for which rebuilding does not lead to change in the checksum. 

\subsection{Redeployment}
Merely changing the image layer's checksum in image's json bypasses the integrity check locally and lets the user run the image smoothly. But this is not without some major concerns. When Docker pushes the image to a remote registry, the integrity test would not pass, because the image will use each layer's id to fetch the same layer id from remote and compare checksum trace. After the code injection, the checksum of local layer vs the remote layer changes; meanwhile, the layer's id remains constant. Since the user cannot change the remote image's content, the checksum bypass strategy does not work. Additionally, if injecting the code changes the content of the layer without changing the ID, another image that is still using the content from the old layer has no choice but to use the new content. To address these concerns, before code injection, we clone the layer in the local registry, so there are two identical layers. Then operate code injection and checksum bypass on one of the layer. When completed, a new Layer ID associated with a new layer will be generated. Then inject the reference of the new layer into image manifest and json to replace the old layer id. Now this image will be accepted by the remote registry as an updated image. 

\section{Performance}
To minimize the time consumed in inter-container communication, we decided to run the experiment on an RT-enabled Linux kernel at a maximum of one container depth as \cite{philip} and \cite{adapt} suggested. To access the performance of the proposed code injection method, we compared the time taken to rebuild an image after changing a source file, between using the original Docker method and our proposed method. The test was conducted in the following four scenarios:

\begin{enumerate}
\item One-line Python project injects 1 line. We base the test image on Alpine and add a simple one-line Python script. This should be the most basic setup for a Python project in Docker. Before rebuilding the image, we append one extra line to the Python script.

\item Complex Python project injects 1000 lines. The test image is based on miniconda3, and a multitude of dependencies will be applied after copying the main Python script. In this case, we append 1000 extra lines prior to rebuild.

\item One-line Java project injects 1 line. In this scenario, for the rebuild, we first add one line to the source code and compile it independent of the docker build process. We then base the test image on java-:8-jdk-alpine and add the compiled code. This should be common for small scale and simple Java project: since there are virtually no dependencies, compiling outside Docker does not risk entangling the environment.

\item Complex Java project injects 1000 lines. We will use Ubuntu as the base image, then install JDK and all the dependency. Next, copy the source code and compile. This approach is usually used in a larger project to achieve better isolation of the environment. Moreover, as similar to the complex Python project, we will add 1000 lines to the source code before rebuilding. Note that when testing with the proposed method, we must not only inject code in the layer containing the source code but also rebuild the layer after it that compiles the source code.
\end{enumerate}

\begin{figure}[htbp]\centerline{\includegraphics[width=\linewidth]{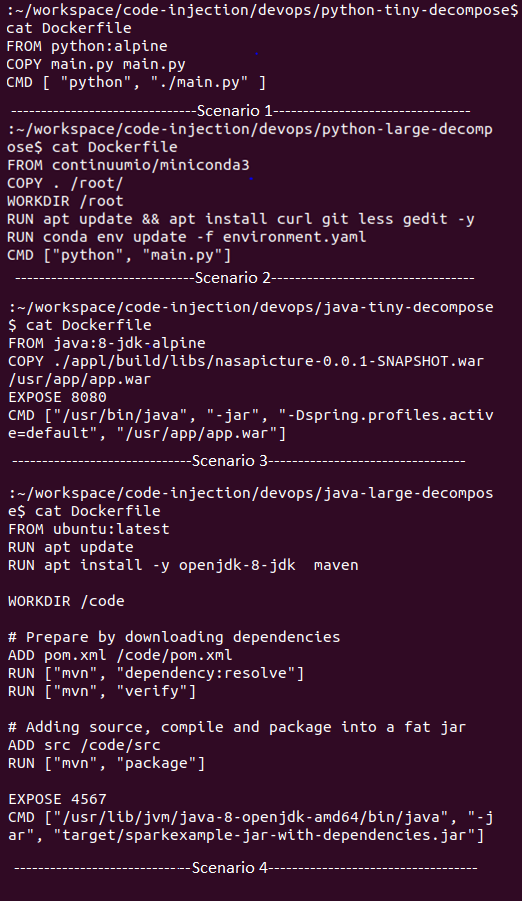}}\caption{Dockerfile of the Four Scenarios}\label{figmanifestjson} \end{figure} 

These scenarios are selected to cover the two extremes: a small project with little changes and a complex project with a lot of changes. Furthermore, the difference between interpreted and compiled languages will be examined. Each scenario was executed for 100 trials and the collected data are presented in Figure 5 and 6.

\begin{figure}[htbp]\centerline{\includegraphics[width=\linewidth]{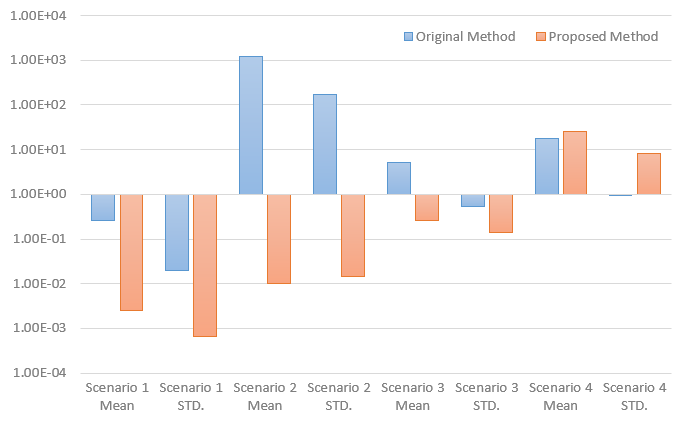}}\caption{Image Rebuilt Time Mean and Standard Deviation}\label{figmanifestjson} \end{figure}

\begin{figure}[htbp]\centerline{\includegraphics[width=\linewidth]{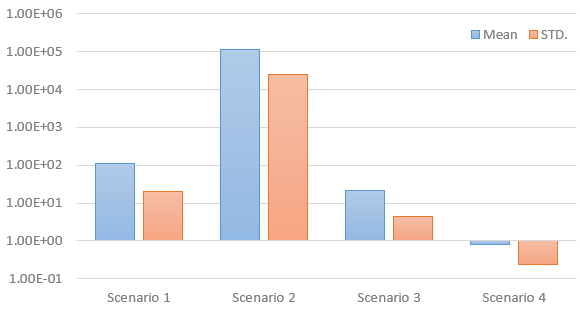}}\caption{Proposed Method Number of Times Faster Than Docker Method}\label{figmanifestjson} \end{figure}

We then perform hypothesis test where the null hypothesis is \(\mu \leq H_0\) . In which \(\mu\) is the real mean for number of times the proposed method is faster and \(H_0\) is the hypothesized mean. We choose to use a low significance level of 0.001 and to employ the \(Z\) distribution. P values are then calculated with the following equation:
\begin{equation}
    P = \phi ( \frac{\mu - H_0}{s/\sqrt{n}} )
\end{equation}

\begin{table}[h!]
\caption{Hypothesis Test Results For Each Scenario}
\label{table:2}
\centering
 \begin{tabular}{||c c c c c||} 
 \hline
  &  Scenario 1 &  Scenario 2 &  Scenario 3 &  Scenario 4 \\ [0.5ex] 
 \hline\hline
 \(H_0\) & 100 & 105000 & 20 & 0.7\\ 
 \hline
  \(P\) & 0.0000026 & 0.0000096 & 0.0001135 & 0.0002309
\\ 
 \hline
\end{tabular}\end{table}

As shown in Table \ref{table:2}, the \(P\) values are lower than the chosen significance level; thus, we reject the null hypothesis. This means that for Python, our proposed code-injection solution could be 100 to more than 100000 times faster than the current rebuilt mechanism Docker used. This number could be even higher for a more complicated project as more time can be saved on eliminating efficiencies in layers fall through and large layer rebuilding over small changes. 

Meanwhile, we see about 20 times improvement in performance in scenario 3. However, scenario 3 is not representative of running the proposed method for compiled languages as it skipped the compilation step in its docker build process. This is not standard for complicated projects seen in industry practices. A more realistic comparison would be scenario 4, in which we see no significant improvement in performance, but in fact a slight degradation. This should be expected because either the proposed method or the original method has to rebuild the layer that compiles the source code. This removes the proposed method's advantage in avoiding layer fall through. Although the proposed method might have saved some time for not rebuilding the layer containing the source code, the performance improvement provided of such is likely outweighed by the extra time introduced in searching and replacing the layer checksum.

\section{Conclusion and Recommendation}
Overall, it can be concluded that for Python, our proposed code-injection method accelerates the rebuild process for at least two magnitudes, and at most five magnitudes times comparing to Docker’s current implementation. We would expect a similar increase in performance for other interpreted languages since the underlying principle of this increase is unchanged: interpreted languages require no compiling; thus, allowing us to avoid layer fall through. 

On the other hand, there is no improvement in performance for Java, or even some decrease in the case of a slightly complicated project commonly seen in industry. We also expect a similar result to be found in other compiled languages as layer fall through cannot be avoided while the improvement to other inefficiencies (uneven layer content distribution and larger layer rebuild for small changes) should be on the same scale comparing to the case for interpreted languages. 

Eventually, we intend to ship this feature either as a Docker plugin or along with other features we developed as a part of a stand-alone containerization tool once it reaches a stable state. All code used in this paper is offered to the community under MIT license for research and reference at \cite{ourrepo}. At this point, using literal injection cannot guarantee integrity for compiled programming languages as compiling to binary code may behave differently than their original program code. Moreover, it is demonstrated that the injection method offers limited acceleration for compiled languages. Lastly, applying the same code injection mechanism, we will proceed to investigate the mechanism of performing multi-layer injection in a forthcoming investigation.

\end{document}